\documentclass[aps,prl,twocolumn,showpacs,nofootinbib,superscriptaddress,groupedaddress]{revtex4-1}
\usepackage{amsmath}
\usepackage{epsfig}
\usepackage{graphicx}
\usepackage{xcolor}
 \usepackage{enumitem}
\usepackage{amssymb}
\usepackage{epstopdf}
\usepackage[utf8]{inputenc}
\usepackage[T1]{fontenc}
\usepackage{ulem}
\usepackage{lineno}
\usepackage{hyperref}
\begin{document}
\title{Stability of dark solitons in a bubble Bose-Einstein condensate}
\author{Raphael Wictky Sallatti$^{1}$}\email{wictkyr@gmail.com}
\author{Lauro Tomio$^{2}$}\email{lauro.tomio@unesp.br}
\author{Dmitry E. Pelinovsky$^{3}$}\email{pelinod@mcmaster.ca}
\author{Arnaldo Gammal$^1$}\email{gammal@if.usp.br}
\affiliation{$^{1}$Instituto de F\'{i}sica, Universidade de S\~{a}o Paulo, 
05508-090 S\~{a}o Paulo, Brazil\\
$^{2}$Instituto de F\'isica Te\'orica, Universidade Estadual Paulista, 
01156-970 S\~ao Paulo, SP, Brazil \\
$^{3}$Department of Mathematics, McMaster University, Hamilton, Ontario, L8S 4K1, Canada}
\date{\today}
\begin{abstract}
The stability of nonlinear waves on curved surfaces is a problem of widespread interest across physics. 
Here, we establish the stability criteria for dark solitons on a spherical Bose-Einstein condensate. 
We demonstrate a sharp instability threshold 
 in the nonlinear parameter,
beyond which solitons 
decay into vortex dipoles via snake instabilities. Analytically and numerically, we prove this decay 
is dictated by a single unstable mode for each angular momentum $m \geq 2$, which is 
a universal mechanism that controls the resulting vortex state.
Unlike in the full three-dimensional case, 
where snake instabilities lead to vortex rings, a dark 
soliton confined to the surface of a bubble can only decay 
into vortex pairs.
\end{abstract}
\maketitle
Bose-Einstein condensed gases in spherical geometries have recently attracted attention due to
experiments performed aboard the International Space Station in a microgravity environment with
ultra-cold gases confined in spherical and/or ellipsoidal surfaces~\cite{2020Aveline, 2022Carollo}. 
These experimental investigations were designed based on previous theoretical work on
shell-like potentials~\cite{2001Zobay,perrin,2004Zobay}.
Ground-based experiments also revealed shell bubbles, by exploring two species~\cite{2022Jia}, 
or intending to observe alternative two-dimensional (2D) closed geometries~\cite{Guo2022}. 
Such investigations have shed new light on the physics of low-dimensional quantum gases, 
especially in closed 2D shells. See Ref.~\cite{2025Dubessy} 
for a recent overview of the present status and perspectives on quantum gases in bubble traps. 
Several interesting problems have been studied in this context, 
considering fundamental physics properties in shell-like structures~\cite{2023Tononi,
2017Padavic,2018Padavic,2019Bereta,2019Tononi,2020Moller}, 
vortex dynamics and stability \cite{2020Padavic,2021Zeng,2021Andriati,2022Caracanhas}, dipole
interaction effects~\cite{2012Adhikari,2020Diniz}, Berezinskii-Kosterlitz-Thouless (BKT) 
transition \cite{2022Tononi}, thermodynamic properties of the gas adiabatic expansion
from filled sphere to hollow one \cite{2021Rhyno}, and to our main interest, some
attention has been paid to the properties of condensate mixtures trapped on a
bubble~\cite{2022Wolf, 2023brito,2022Sturmer}. Furthermore, the study of dimers
on a spherical surface shows that the dimers are squeezed in the
direction orthogonal to the center of mass motion, qualitatively changing its
geometry, from 2D to one-dimensional (1D), leading to two-soliton
motion on a bubble surface~\cite{2025Tononi}.

In the context of the properties of Bose-Einstein condensates (BECs) on spherical
surfaces, a topic of particular interest concerns the existence, stability, and
propagation of dark solitons~\cite{2016Wang}. Bubble BECs are realized when atoms are confined
to a thin spherical shell, creating unique curvature effects that influence the dynamics of solitons. 
Dark solitons are shape-localized propagating waves manifested by depressions
in the condensate density. 
On a bubble surface, geometric effects can alter the instability dynamics of dark 
solitons, as compared to their propagation on flat surfaces. 
The curved surface is expected to affect their dynamics, propagation speed, and stability.
Unlike in flat quasi-1D or quasi-2D BECs, where dark solitons are susceptible to snake 
instability~\cite{anglin2021}, decaying into vortices, the closed topology of a bubble is 
expected to alter the decay pathways, with the curvature introducing geometric constraints 
affecting the stability dynamics. Therefore, the soliton can experience strain 
due to the curvature, leading to contraction or expansion depending on the interactions.

The instability of a dark soliton stripe on a sphere follows a procedure similar 
to that in flat space. However, a sphere has no edges, such that 
the curvature influences perturbations that would be symmetric in a flat system.
The most critical difference in a bubble refers to the fragmentation into vortices, as the
resulting vortices are not free to move arbitrarily. The snake instabilities on
a flat plane produce vortex-antivortex pairs that can move apart from each other and 
eventually are ejected from the high-density region.
On a sphere, there is no edge to eject vortices as in flat space, with 
vortex-antivortex pairs remaining trapped on the surface, leading to complex dynamics like 
vortex-antivortex annihilation or formation of stable stationary patterns.
The profound topological constraint implies that a single vortex is not allowed on a sphere, as
vortices must exist in pairs of opposite circulation whose charges sum to zero.

Dark solitons in a flat BEC are known to oscillate and decay due to quantum fluctuations 
or dissipation. In quasi-1D traps, dark solitons can be metastable and decay due to transverse 
instabilities~\cite{1999Burger,2002Muryshev}. Using harmonic traps, they were studied 
in~\cite{2000Busch,2000Feder}, both analytically and computationally,
where it was found that the large-amplitude field modulations at a frequency resonant with 
the energy of a dark soliton give rise to a state with multiple vortices. The stability 
spectrum of the dark soliton contains complex frequencies, which disappear for sufficiently 
small numbers of atoms or for a large transverse confinement. The relationship between these 
complex modes and the snake instability was investigated numerically by real-time 
propagation, see also ~\cite{2001Dutton}.  In ring geometries, dark solitons in BEC systems, 
introduced in \cite{1994Kivshar}, were further investigated in~\cite{2003Theocharis}.
Finally, it is also important to note the growing interest 
in dark-soliton dynamics driven by the application of holographic 
methods to spherical superfluids, as verified in~\cite{2025Gao,2025Gao-pp}.

Two-component BECs were considered in~\cite{2001Anderson}, where the dark soliton exists 
in one of the condensate components and the soliton nodal plane is filled with the second 
component. The filled solitons are stable for hundreds of milliseconds. 
By selectively removing the filling, one can make the soliton more susceptible to 
dynamical instabilities. For a condensate in a spherically symmetric potential, these 
instabilities cause the dark soliton to decay into stable vortex rings.
By studying the oscillations and interactions of dark and dark-bright solitons, it was also
shown in ~\cite{2008Becker} that the stability of solitons can be controlled in BECs, 
via confinement. 

The aim of the present study concerns on the spectral stability of 
dark solitons under discrete $m-$angular modes within a BEC confined on the surface 
of a bubble in the approximation of a 2D spherical hollow shell~\cite{2021Andriati}. 
We show the occurrence of exactly one unstable mode for each $m \geq 2$,
which induces a snake-like excitation that breaks the condensate into pieces,
as observed in~\cite{2021Andriati}. As verified numerically, an $m-$dominated unstable 
mode is followed by the creation of $m$ vortex-antivortex pairs, in agreement with our
analytical predictions. 

Next, we first present the theoretical model 
for an atomic BEC confined to the surface of a perfect sphere, named a bubble. 
Dark solitons, together with the asymptotic approximations in the limit of small and large 
chemical potentials, are obtained analytically and numerically, supported by 
Supplemental Material (SM)~\cite{SP}. We then report the 
analytical results of the corresponding stability analysis, demonstrating they are in 
excellent agreement with numerical observations. Finally, we report outcomes of the 
nonlinear dynamics of the system, from which it follows that the instabilities result 
in the formation of vortex dipoles on the dark soliton condensate. 

{\it Mathematical model --- }
The present study is performed by assuming the condensate is trapped on the surface of 
a rigid spherical shell, aiming to mimic the cold-atom bubble experiments that are currently 
being performed in microgravity environments. In this approach, the system can be studied by 
considering a reduction of the three-dimensional (3D) Gross-Pitaevskii equation (GPE) to a corresponding 2D 
system described by spherical coordinates $\theta \in [0,\pi]$ and $\phi \in [0,2\pi]$, 
with fixed radius $R > 0$, which is also assumed to be the space unit. In the dimensional 
reduction, $\delta R$ is also taken as being the radial thickness of the spherical bubble.
The two-dimensional approximation is reasonable since excitations in the radial direction are
inaccessible due to the large amount of energy required for them. This is true when the 
thickness $\delta R$ of the 3D spherical shell is very small compared to its radius $R$, 
that is, $\delta R \ll R$, as discussed in Refs.\cite{2021Andriati,2023brito}. We also stress 
that our main concern is the {\it dynamic stability} of the dark soliton. We do not take 
into account its {\it energetic instability}, which could be triggered by a thermal cloud 
that is neglected here. The condensate can be described in the mean-field approach as a 
system of two coupled GPEs~\cite{1999Dalfovo,2016Stringari} with the nonlinear two-body 
parameter $\overline{g} \equiv {4\pi \hbar^2{a_s}N }/{M}$,
where $N$ is  the number of atoms with mass $M$, and $a_s$ is the atom-atom $s$-wave scattering 
length, that we assume to be positive in our approach for a generally stable condensate. 
Given $R$ as the length unit, with the time in units of $MR^2/(2\hbar)$, the dimensionless 2D GPE for 
the wave function $\psi\equiv\psi(\theta,\phi,t)$, normalized to one, is written as
{\small\begin{eqnarray}
{\rm i}{\partial_t \psi}
&=&- 
\Delta_{\rm 2D} \psi + g {\left| \psi\right|}^{2}\psi\label{eq01}\\
&\equiv&-\left[\frac{1}{\sin\theta}{\partial_\theta}\left(\sin\theta{\partial_\theta}\right)+
    \frac{1}{\sin^{2}\theta}{\partial^2_\phi}\right]
\psi + g {\left| \psi\right|}^{2}\psi,\nonumber
\end{eqnarray}
}where $g \equiv{4 \sqrt{2\pi}a_s N}/{\delta R}$.
With $\ell$ and $m$ being the quantum numbers related to the angular variables 
$\phi$ and $\theta$, a given state $\psi$ is given by $\psi_{\ell m}$. 
By separating the $m$-angular mode, with $\psi_{\ell m}\equiv 
e^{{\rm i}m\phi}\Theta_{\ell m}(\theta)$, the above defined 
$\Delta_{\rm 2D}$ is reduced to the $1D$ operators $\Delta_m$, leading to 
the Laplace equation
{\small\begin{equation}
\label{m-Delta}
\!\!-\Delta_m\Theta_{\ell m} = 
-\left[\frac{d^2}{d\theta^2} + \cot \theta\;\frac{d}{d \theta} - \frac{m^2}{\sin^2\theta} 
\right] \Theta_{\ell m} = \lambda_\ell \Theta_{\ell m}.
\end{equation}
}Recall that bounded solutions of this
equation 
exist if and only if 
$\lambda_\ell=\ell(\ell+1)$, with $\ell \in \mathbb{N}_0 = \{0,1,2,\dots\}$, and 
$-\ell \leq m \leq \ell$. 
They can be expressed by the associate Legendre polynomials, 
with $\Theta_{\ell m}(\theta)=P_\ell^m(\theta)$. 

{\it Existence of dark solitons ---}  The existence of dark solitons on a bubble
(dark ring solitons), previously investigated in 
\cite{1994Kivshar,2003Theocharis}, can be associated with the stationary solution of 
the GPE,
{\small\begin{eqnarray} 
\label{0-Delta}
-\left( 
\frac{d^2}{d\theta^2} + \cot \theta\;\frac{d}{d \theta}\right)
f(\theta) &+& \varepsilon|f(\theta)|^2f(\theta)=\mu f(\theta),
\end{eqnarray}
}where $\varepsilon=g/(2\pi)$ is the magnitude of the  
defocusing nonlinearity, and 
$\mu$ is the chemical potential.
To obtain \eqref{0-Delta} from \eqref{eq01}, we assume
\begin{equation}
\psi_s(\theta,\phi, t) = \frac{f(\theta)}{\sqrt{2 \pi}} e^{-i \mu t},    
 \;\;   \int_0^\pi  \sin{\theta} |f(\theta)|^2 \text{d}\theta = 1.
\label{eq02}
\end{equation} 
Eq.~\eqref{0-Delta} admits the conservation of flux, 
\begin{equation}
\label{flux}
J = \sin \theta \left[ \bar{f}(\theta) \frac{d f(\theta)}{d\theta} - \frac{d\bar{f}(\theta)}{d\theta} 
f(\theta) \right], \quad 0 < \theta < \pi, 
\end{equation}
where $f(\theta)$, $df(\theta)/d\theta$ are bounded at zeros of $\sin \theta$. 
Hence, we have $J = 0$ and, without loss of generality, 
we can consider a real-valued function $f(\theta) :[0,\pi] \to \mathbb{R}$.
The dark soliton profile is defined by a density that decreases monotonically,
$|f(\theta)|^2$ on $[0,\pi]$, with zero at $\theta = \frac{\pi}{2}$. 
Given the normalization 
\eqref{eq02}, $\mu$ is defined by $\varepsilon$, as
\begin{equation}
\label{mu}
\mu = \int_0^\pi \sin{\theta} \left( \left|\frac{d f(\theta)}{d\theta}\right|^2 + 
\varepsilon |f(\theta)|^4 \right) \text{d}\theta. 
\end{equation}	
A constant solution of \eqref{0-Delta} exists for 
every $\varepsilon > 0$ in the form:
$f(\theta) = {1}/{\sqrt{2}}$, with
$\mu = \varepsilon/{2}.$
A dark soliton is obtained by the small $\varepsilon$ limiting solution of \eqref{0-Delta}:
{\small\begin{equation}
f(\theta) = \sqrt{\frac{3}{2}} \cos \theta + \varepsilon f_1(\theta) +  \mathcal{O}(\varepsilon^2), \;\;
\mu = 2 + \frac{9}{10} \varepsilon + \mathcal{O}(\varepsilon^2).
\label{dark-soliton}
\end{equation} 
}As shown in~\cite{SP}, $f_1(\theta)=\frac{3\sqrt{3}}{100\sqrt{2}}\left[3\cos\theta-5\cos^3\theta\right]$
is uniquely defined and proportional to $P_3(\cos\theta)$.
As the main term of \eqref{dark-soliton} is the normalized 
$P_1(\cos\theta)$, the orthogonality of the Legendre polynomials imply that the 
normalization \eqref{eq02} for small $\varepsilon$ is restricted to $\mathcal{O}(\varepsilon^2)$.

In the other extreme, 
with $\varepsilon \to \infty$, 
the dark soliton profile becomes concentrated near the equator ($\theta = \pi/2$), 
according to the exact solution in the flat space:
{\small\begin{eqnarray}
-\frac{d^2}{d\theta^2}f_{\infty}(\theta)&+& \varepsilon f_{\infty}^3(\theta) = 
\mu_{\infty}(\varepsilon) f_{\infty}(\theta),
\label{dark-soliton-lim}\\
f_{\infty}(\theta) &=& \frac{1}{\sqrt{2}} \tanh\left[ \frac{\sqrt{\varepsilon}}{2} \left( \frac{\pi}{2} 
- \theta \right)\right], \; \mu_{\infty}(\varepsilon) = \frac{\varepsilon}{2},\nonumber 
\end{eqnarray}
}where $f_{\infty}(\theta)$ remains normallized to one only in the exact asymptotic
limit. With the convenient variable change
$z\equiv  \frac{\sqrt{\varepsilon}}{2} \left( \frac{\pi}{2} - \theta \right)$,
this solution can be incorporated into the asymptotic expansion. As shown in \cite{SP},
where $f(\theta)\equiv g(z)$, $ f_{\infty}(\theta)\equiv g_0(z)$, and $f_\infty^{(1)}(\theta)\equiv g_1(z)$:
{\small\begin{equation}
\label{dark-soliton-large}
f(\theta)= f_{\infty}(\theta)+\frac{f_\infty^{(1)}(\theta)}{\sqrt\varepsilon} 
+\mathcal{O}\left(\frac{1}{\varepsilon}\right),
\mu = \frac{\varepsilon}{2} + \sqrt{\varepsilon}+
\mathcal{O}\left(1\right),
\end{equation}
}with $f_{\infty}^{(1)}(\theta)$ uniquely defined by $g_1(z)=\frac{d}{dz}[zg_0(z)]$.

\begin{figure}[h]
 \hspace{-0.4cm}           \includegraphics[scale=0.16]{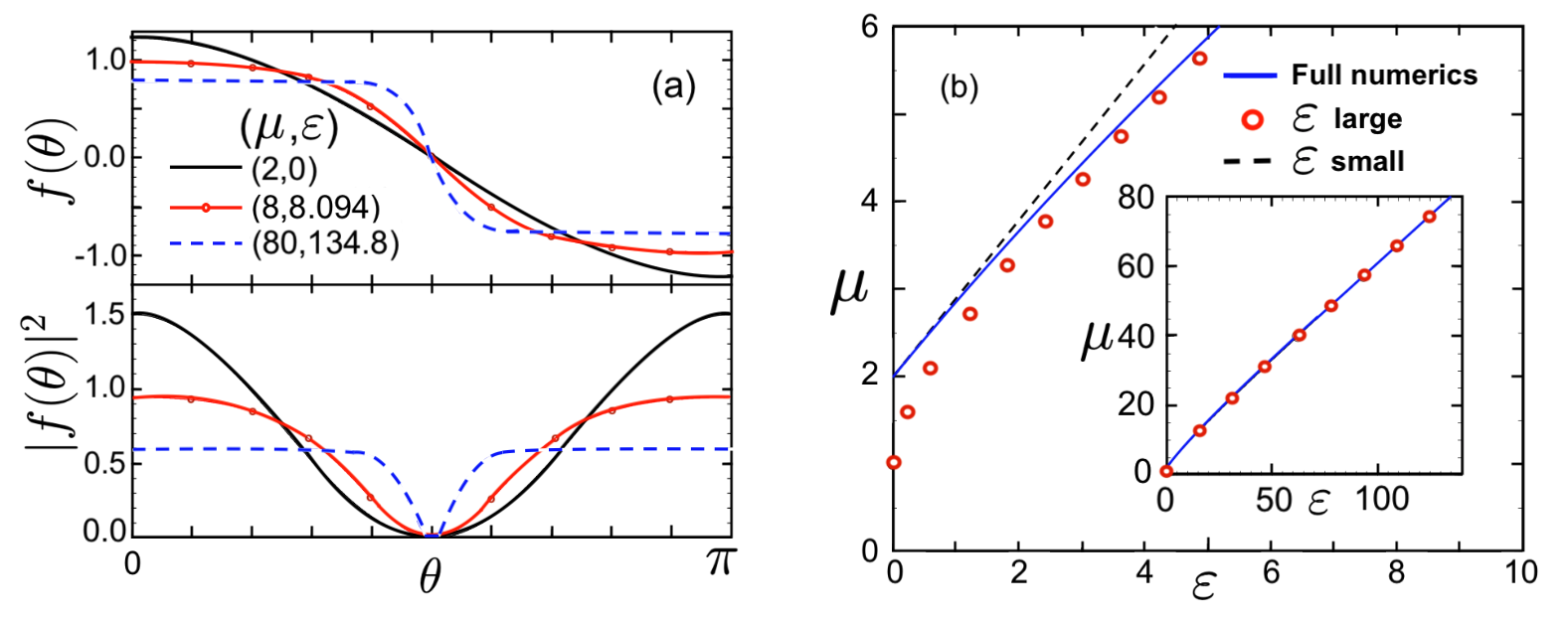}
\vspace{-0.4cm}
\caption{(Color on-line)
In (a), we show profiles of dark solitons, $f(\theta)$ (upper panel),
with corresponding densities $|f(\theta)|^2$ (lower panel), as function
of $\theta$, for three different values of $\mu$, with
$\varepsilon$ obtained numerically by using the shooting method.
In (b), the chemical potential $\mu$ is presented as a function of 
$\varepsilon$ for dark solitons. Solid lines correspond to the 
numerical data, with dashed and circles referring to  
$\mu=2+\frac{9}{10}\varepsilon$ and $\mu=\frac{\varepsilon}{2}+\sqrt{\varepsilon}+1$, 
respectively, in agreement with \eqref{dark-soliton} 
and \eqref{dark-soliton-large}.}
\label{fig01}
\end{figure}

In Fig.~\ref{fig01}(a) we present profiles of the dark soliton obtained numerically, 
by using the shooting method described in \cite{SP}. The profile $f(\theta)$ is 
monotonically decreasing with the only zero at $\theta=\pi/2$ (the equator). 
It is close to $\sim \cos \theta$ as $\varepsilon \to 0$ according to \eqref{dark-soliton} 
and close to $f_{\infty}(\theta)$ as $\varepsilon \to \infty$ according to 
\eqref{dark-soliton-large}. 
We confirm the asymptotic predictions by plotting $\mu$ versus $\varepsilon$ 
in Fig.~\ref{fig01}(b), which shows a good agreement between the theory and numerical 
results.

{\it Stability of dark solitons --- } Linear stability analysis~\cite{GP25} 
can be performed to verify the stability of dark solitons. In this context, 
the well-known Bogoliubov-de Gennes (BdG) method was considered 
in~\cite{2021Andriati}. Within this approach, small amplitude oscillations 
are assumed around the stationary solution \eqref{eq02}, which is redefined as  
{\small\begin{equation}
\label{decomposition}
\psi(\theta,\phi,t) = \psi_s(\theta,\phi,t)+\frac{e^{-{\rm i} \mu t}}{\sqrt{2\pi}} 
\left[u(\theta,\phi,t) + {\rm i} v(\theta,\phi,t) \right], 
\end{equation}
}with $u$ and $v$ being real-valued. From Eqs.~\eqref{eq01}, \eqref{eq02}, and 
\eqref{decomposition}, we obtain the linearized equation of motion, which leads to 
the following coupled equations for $u$ and $v$:
\begin{eqnarray}\label{nls-lin}
\left\{ \begin{array}{c}
\partial_t u = -\Delta_{\rm 2D} v + \varepsilon f^2(\theta) v - \mu v, \\
-\partial_t v = -\Delta_{\rm 2D} u + 3 \varepsilon f^2(\theta) u - \mu u,
\end{array} 
\right.
\end{eqnarray}
where the chemical potential $\mu$ is a function 
of $\varepsilon=g/(2\pi)$, given by \eqref{mu}. 
Performing the separation of variables for $m$-angular modes in $\phi$, with
$(u,v)$ given by the eigenvectors $(\hat{u}_m,\hat{v}_m)$ and
eigenvalue $\omega$,
\begin{equation}
\label{separation}
u = \hat{u}_m(\theta) e^{{\rm i} (m \phi + \omega t)}, \quad 
v = \hat{v}_m(\theta) e^{{\rm i} (m \phi + \omega t)}
,\end{equation}
we define the spectral stability problem. 
For each $m$-mode separately, we have the coupled system 
\begin{equation}\label{nls-spectrum}
\left\{ \begin{array}{l} 
\omega \hat{u}_m = L^-_m \hat{v}_m, \quad L^-_m = -\Delta_m  + 
\varepsilon f(\theta)^2 - \mu, \\
\omega \hat{v}_m = L^+_m \hat{u}_m, \quad L^+_m = -\Delta_m  + 
3\varepsilon f(\theta)^2 - \mu, 
\end{array} 
\right.
\end{equation}
with $\Delta_m$ given in \eqref{m-Delta}. 

The dark soliton is spectrally stable with respect to the $m$-angular 
mode if the corresponding imaginary part of $\omega$ is zero, 
for all eigenvalues of \eqref{nls-spectrum}.
Let us consider, for each mode $m$, the corresponding $\omega_m$.
As we know from the general theory \cite{GP25}, all eigenvalues 
$\omega_m$ of the spectral stability problem \eqref{nls-spectrum} 
are real if the linear operators $L^-_m$ and $L^+_m$ are positive. 
These linear operators enjoy two comparative relations:
\begin{eqnarray}
&L_m^+ - L_m^- = 2 \varepsilon f^2(\theta) \geq 0 & 
\label{compar-1}\\
{\rm and\hskip 1cm}&&\nonumber\\
&L_{m+1}^{\pm} - L_m^{\pm} = \frac{2m+1}{\sin^2\theta} \geq 0.&
\label{compar-2}
\end{eqnarray}
All the eigenvalues of $L_m^\pm$ are strictly positive 
for $m \geq 2$ and small $\varepsilon > 0$, as shown in \cite{SP}.
With $L_m^+ > L_m^-$  for all $\theta$, which are in the closed
interval $0\le\theta\le\pi$ (excluding $\theta={\frac{\pi}{2}} $), as well as for  
$L_{m+1}^{\pm} > L_m^{\pm}$ with all $\theta$ in the open interval $0<\theta<\pi $, 
the smallest eigenvalue of $L_m^-$ is smaller than the smallest one of $L_m^+$; 
with the smallest eigenvalue of $L_m^{\pm}$ being smaller than the smallest one of
$L_{m+1}^{\pm}$. The same holds for the second smallest 
eigenvalues of the same operators in the subspace of odd functions about 
$\theta = \frac{\pi}{2}$. 
Hence, the dark soliton is stable with respect to the 
$m$-angular mode for small $\varepsilon$. However, the 
lowest eigenvalue of $L_m^-$, defined as $\omega^-_m$, can cross zero
and become negative for sufficiently large $\varepsilon$,
triggering the snaking instability of the dark solitons.

Next, by considering \eqref{compar-2}, it is demonstrated in \cite{SP} that 
the smallest eigenvalue of $L_{m+1}^-$ crosses zero for values of $\varepsilon$ 
larger than the smallest eigenvalue of $L_m^-$ for any given $m \geq 2$, which 
always occurs for large values of $m$. 
This implies that, for 
every $m \geq 2$, there exists a $\varepsilon_m > 0$ satisfying 
$\varepsilon_m < \varepsilon_{m+1}$, such that $L_m^-$ has a simple 
negative eigenvalue for $\varepsilon > \varepsilon_m$. This 
further clarifies that the spectral stability problem \eqref{nls-spectrum} 
provides a real unstable eigenvalue $\omega_m$ for 
$\varepsilon > \varepsilon_m$.
As shown in Table~\ref{tab1}, the asymptotic expression, derived in~\cite{SP} 
for $m\gg 2$,
%{\small 
\begin{equation}
\label{asympt-eps-large}
\varepsilon_{m}^{th} \approx 4 m (m-1),
\end{equation}
provides a good approximation to predict the threshold onset of 
instability, considering {\it all values} of $m\ge 2$. The 
$\mathcal{O}(1)$ discrepancy between \eqref{asympt-eps-large} and the 
numerical data represents less than 5\% even for $m=2$.
\begin{table}[!h]
    \caption{For each angular mode $m$, the threshold values [where ${\rm Im}(\omega_m)=0$] 
    of $\varepsilon$ are given, with $\varepsilon_m$ being the exact numerical results and 
    $\varepsilon_m^{th}$ given by the analytical approximation \eqref{asympt-eps-large}. 
    The respective values of $\mu$ are also presented.     }
    \label{tab1}
    \centering
        \begin{tabular}{c|cccccc}
\hline\hline
           $m$      &  2  &  3   &  4   &    5 & 6    &  7 \\ 
           \hline\hline
$\varepsilon_m$     &8.367&24.402&48.416&80.420&120.420&168.420\\
$\varepsilon_m^{th}$&8    &24   &48     &80    &120   &168\\
$\mu$               &8.182&18.202&32.208  &50.210&72.210&98.210\\
\hline\hline
    \end{tabular}
\end{table}

\begin{figure}[h]
\centering \hskip -0.5cm
\includegraphics[scale=0.18]{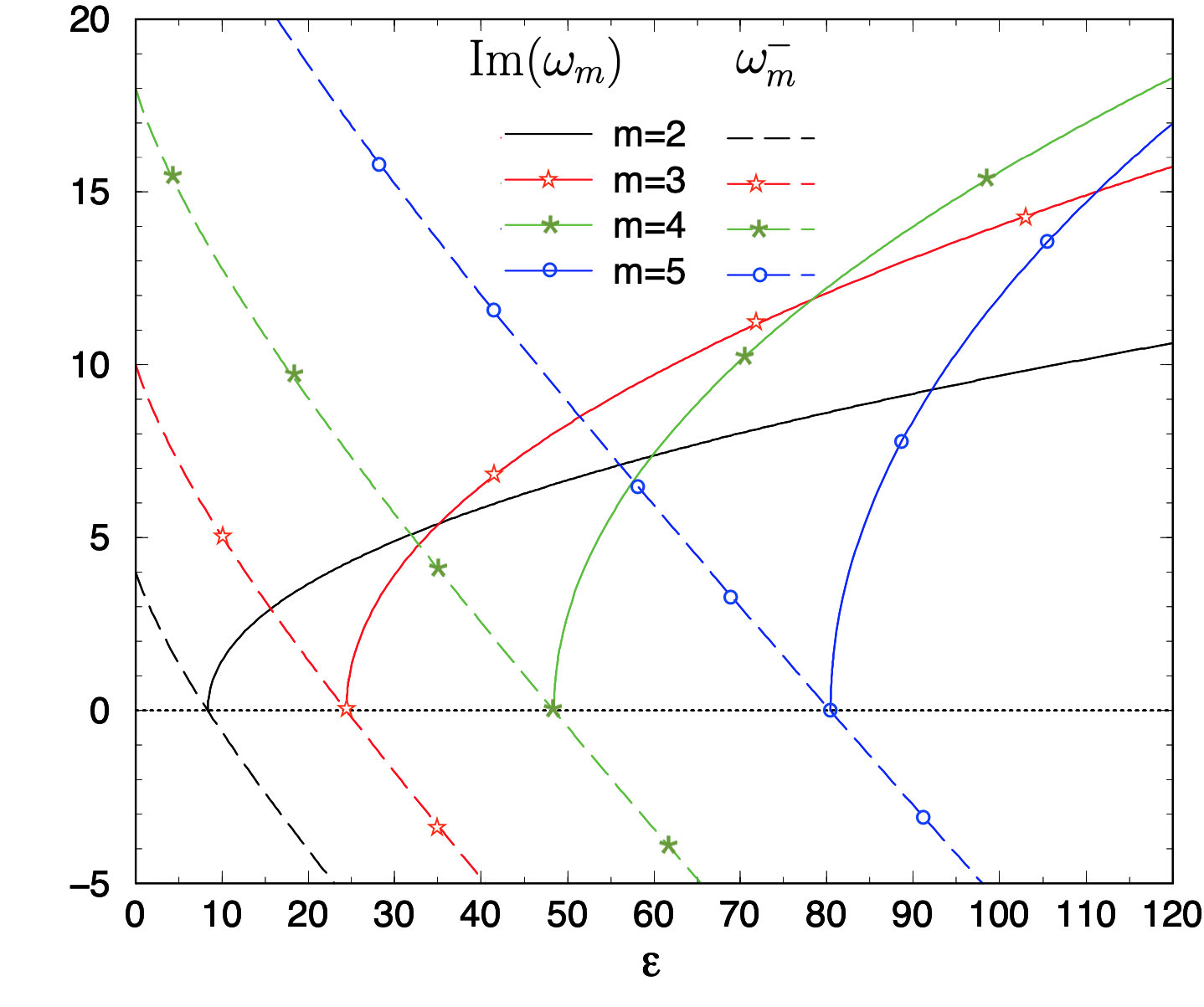}
\vspace{-0.4cm}
\caption{(Color on-line)
By varying $\varepsilon$, we show for different $m$-angular modes 
(indicated by the symbols and lines) that the 
imaginary part of the eigenvalues $\omega$ (positive defined), 
obtained from \eqref{nls-spectrum} (solid lines), start increasing from 
zero when the corresponding lowest eigenvalue $\omega^-_m$ of $L_m^-$ 
(dashed lines) becomes negative. For a given $\varepsilon$, the
dominant unstable mode $m$ is provided by the largest 
${\rm Im}(\omega_m)$.}
\label{fig02}
\end{figure}

The analytical predictions are confirmed numerically, as shown in  
Fig.~\ref{fig02} for the angular modes with $m = $ 2, 3, 4, and 5. 
The onset of instability in the spectral problem \eqref{nls-spectrum} 
corresponds to the zero eigenvalues of the linear operator $L_m^-$,
with the imaginary part of the frequencies dynamically characterizing the 
instability. The dark soliton { 
starts becoming unstable when $\varepsilon \gtrsim 8.37$, with $m=2$ being the initially dominant mode. 
By defining the lower limits of each dominant modes by $\varepsilon_{m}^{(0)}$, when
${\rm Im}(\omega_{m-1})={\rm Im}(\omega_m)$, 
the $m=2$ mode remains dominant for larger $\varepsilon$, until  $\varepsilon_{3}^{(0)}\approx 35$
[when $\rm{Im}(\omega_2)=\rm{Im}(\omega_3)$], where the mode $m=3$ starts dominating the instability.
Next, we have $\varepsilon_{4}^{(0)}\approx 78$ for the mode $m=4$; and 
$\varepsilon_{5}^{(0)}\approx 140$ for $m=5$ (already outside the range shown in Fig.~\ref{fig02}).
}
As represented in Fig.~\ref{fig02}, for a given
$\varepsilon$, the largest imaginary eigenvalue characterizes the dominant 
unstable mode. If a given $\varepsilon$ corresponds to a dominant mode 
$m \geq 2$, then the dark soliton is affected by snake instability and
tends to break into $m$ vortex-antivortex pairs.

\begin{figure}[h]
    \centering
\includegraphics[scale=0.22]{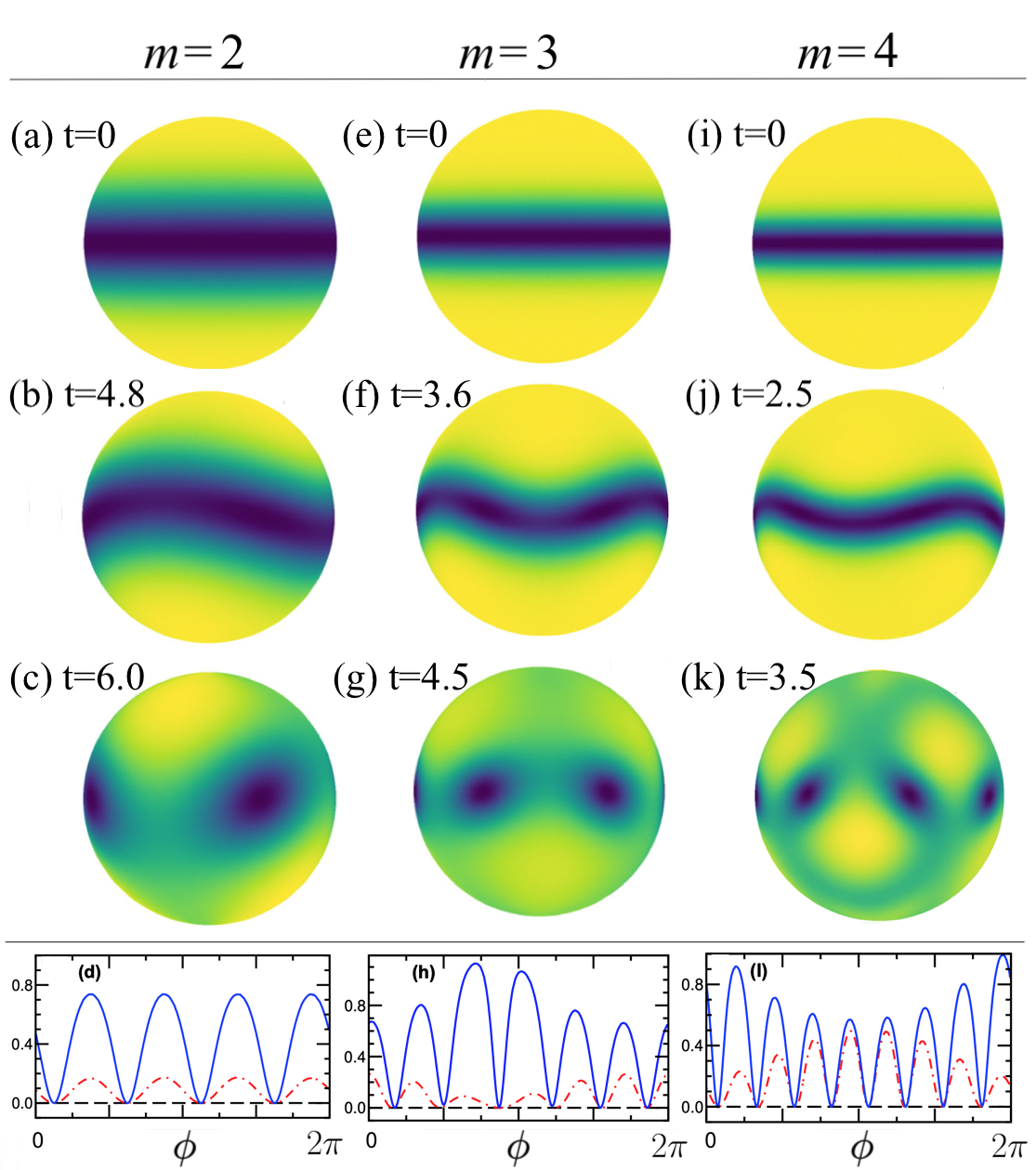}
\vspace{-0.6cm}
\caption{(Color on-line) 
Dark-soliton dynamics, for given instants $t$, with dominant instability modes 
$m=2$ (left panels, $\varepsilon=20$),
$m=3$ (center panels, $\varepsilon=50$), and 
$m=4$ (right panels, $\varepsilon=100$), 
shown by the densities $|\psi|^2$. 
In 3D graphics, the darker the region, the less dense it is.
At the equator, with $t$ corresponding to the upper panels in the same column, 
$|\psi(\frac{\pi}{2},\phi)|^2$ are shown in (d), (h), and (l). 
The bottom dashed lines are for $t=0$; the red dotted-dashed, when snake
instabilities occur; and the solid-blue lines, after the breakup in vortex pairs.}
\label{fig03}
\end{figure}

Regarding the spectral stability of dark solitons for $m = 1$, we demonstrate in 
\cite{SP} that the Eq.~\eqref{nls-spectrum} has only real eigenvalues $\omega$, 
with no instability bifurcations. Finally, for $m = 0$, it is also shown in
\cite{SP} that there exists a single pair of eigenvalues $\omega$ of negative
energy, which are smaller than all other pairs of eigenvalues $\omega$
of positive energy for small values of $\varepsilon$. These pairs can coalesce
hypothetically for large values of $\varepsilon$, triggering another instability
bifurcation. However, we have no supporting numerical evidence for the occurrence 
of such instability.

In Fig.~\ref{fig03} we illustrate the dynamical  
instability of the dark soliton in a sphere for three different dominant 
modes: $m=2$, $m=3$, and $m=4$. The three choices are taken respectively at
$\varepsilon=$ 20, 50, and 100 [See Fig.~\ref{fig02} for the given data].
At $t=0$, we observe the dark-soliton positions at the equator, with their 
respective widths being smaller for larger values of $\varepsilon$. Next, 
in the time evolution, we can verify the onset of snake instabilities 
provoking the breakup of such dark solitons in vortex-antivortex pairs.
As seen in panels (c) and (d), for $m=2$, two pairs of vortices are formed; in
(g) and (h), for $m=3$, we observe the formation of three pairs of vortices; 
and, in panels (k) and (l), for $m=4$, we have four pairs of vortices. 
The widths of the dark solitons, as well as the healing lengths of the 
generated vortices (after the breakup of the snake instabilities),
correspond to $\sim {1}/{\sqrt{\varepsilon}} \approx 0.22$
($m=2$), 0.14 ($m=3$), and 0.10 ($m=4$).

The topology of the sphere implies that, on the onset of instability,
the vortex number change must be $+2$, due to Poincar\'e-Hopf theorem, 
which states that any continuous tangent vector field on a sphere must 
have at least one point where it is zero. For a superfluid, this implies 
no single vortex on a sphere. They must exist in pairs of opposite 
circulation (vortex-antivortex pairs) whose charges sum to zero.

The numerical approach considered in the 
dark-soliton dynamics is carried out first by obtaining the stationary solutions 
using the shooting method \cite{quinney,gammal1999}. The time evolution employs 
the combined spectral method together with the finite difference method introduced 
in \cite{2021Andriati}. For the full numerical calculations of the GPE~\eqref{eq01}, 
we have assumed a time-step $\delta t=10^{-5}$, with spatial grids in $\theta$ and
$\phi$ directions of sizes $256\times256$.  The respective step sizes were
$\delta\theta=\pi/256\approx0.0123$ and 
$\delta\phi=2\pi/256\approx0.0245$. Details of the numerical methods 
are given in \cite{SP}.

{\it Conclusions ---} 
In the present letter, we have reported results obtained numerically and
analytically for the dynamical stability of dark solitons in a single 
atomic BEC trapped on the surface of a rigid spherical bubble with radius
$R$ and thickness $\delta R$. 
All the numerical and analytical results rely on a single parameter 
$\varepsilon$, which encompasses all information on the repulsive two-body 
interaction, the number of atoms, and the thickness of the bubble surface.
The results of our analysis, in terms of the azimuthal excitation angular 
momentum modes $m$, should be useful when considering possible experimental 
realizations. 
We expect these results to impact investigations of BEC systems in spherical 
geometries, which are being performed aboard the
international space station~\cite{2022Carollo}, as well as in shell 
condensates made with atomic mixtures~\cite{2022Jia}. 
Based on the spectral stability analysis, our study in the sphere shows that 
for the interaction parameter $\varepsilon \lesssim 8.37$, dark solitons are 
stable for all $m-$angular mode excitations. For $\epsilon \gtrsim 8.37$, 
modes  with $m\ge 2$ are excited, where the dominant $m-$angular mode induces 
snake-like instabilities that cause the soliton to break up with the production 
of $m-$ vortex pairs.\\
{ In closing, we draw attention to the key findings reported by Anderson et al.~\cite{2001Anderson},
who observed the formation of stable vortex rings resulting from the decay of a dark soliton 
formed in one component of a two-component Bose–Einstein condensate confined within a 
spherically symmetric potential.
The dark soliton produced in one component (once the other component is removed), 
becomes susceptible to dynamical snake instabilities which decay into stable vortex rings. 
Our present results show that this is not achievable within a spherical 2D framework, where 
the condensate is confined to a bubble skin.
To form closed loops in the core—identified as vortex rings—the third dimension is necessary.
Further, we are highlighting that by reducing the 3D system to a quasi-2D geometry within a bubble, 
the vortex rings that typically arise from the decay of dark solitons are replaced by coupled 
vortices localized at the surface. While this outcome may be anticipated, we present a clear 
and specific rule governing the snake instability decay across different instability modes, 
as illustrated in Fig. ~\ref{fig03}.
This finding reveals an important aspect that paves the way for further exploration, particularly 
using experimental setups like the one reported in ~\cite{2001Anderson}.\\
}
In summary, 
for dark solitons on a sphere, we are highlighting the occurrence of a universal mechanism that controls 
the resulting vortex state, based on purely analytical studies compared with exact computational results.
Our predictions for a single BEC in a sphere should provide valuable information for ongoing 
experimental investigations.

{\it Acknowledgements ---}
We acknowledge partial support from Fundação de Amparo à Pesquisa do Estado de São Paulo 
(Procs.~2024/01533-7), Conselho Nacional de Desenvolvimento Científico e 
Tecnológico [Procs. 303263/2025-3 (LT) and 306219/2022-0 (AG)], and 
Coordenação de Aperfeiçoamento de Pessoal de Nível Superior (RWS).

\renewcommand\thefigure{\thesection S\arabic{figure}}
\renewcommand\theequation{\thesection S\arabic{equation}}
\setcounter{figure}{0}
\setcounter{equation}{0}
%\appendix*{Supplemental Material}
\section*{Supplemental Material}
This supplemental material provides analytical and numerical details to corroborate the 
main results obtained in a study on the stability of dark solitons in a Bose-Einstein condensate 
trapped on the two-dimensional surface of a spherical bubble.
\renewcommand\thefigure{\thesection S\arabic{figure}}
\renewcommand\theequation{\thesection S\arabic{equation}}
\setcounter{page}{1}
\setcounter{figure}{0}
\setcounter{equation}{0}
\section{Dark solitons for small $\varepsilon$}
Let us consider the profile $f(\theta) : [0,\pi] \to \mathbb{R}$, bounded at the end points 
$[\theta = 0, \pi]$, defined from the nonlinear differential equation,
\begin{equation}
\label{ode}
-f''(\theta) - \cot \theta f'(\theta) +\varepsilon f^3(\theta)=\mu f(\theta), 
\end{equation}
where we use primes for derivatives, with $\varepsilon > 0$ being a small parameter for the 
defocusing nonlinearity. By imposing the normalization constraint, 
\begin{equation}
\label{normalization}
\int_0^{\pi} d\theta\sin\theta |f(\theta)|^2 = 1,
\end{equation}
there exists a countable set of solutions $\{ \mu_{\ell}\}_{\ell \in \mathbb{N}_0}$, 
uniquely parameterized by $\varepsilon > 0$, which bifurcate from the linear modes 
$\Theta_{\ell}(\theta) = P_{\ell}(\cos \theta)$ 
of the Laplace equation 
\begin{equation}
\label{Laplace}
- \left( \frac{d^2}{d\theta^2} - \cot \theta \frac{d}{d\theta} \right)
\Theta_{\ell}=
\ell (\ell + 1) \Theta_{\ell},
\end{equation} 
where $P_{\ell}(\cos\theta)$ is the $\ell$-degree
Legendre polynomial. Therefore, 
for small $\varepsilon$, a solution of \eqref{ode} can be written as
\begin{equation}
\label{modes}
f(\theta) = \frac{P_{\ell}(\cos \theta)}{\left[ 
\int_0^{\pi}d\theta\sin\theta|P_{\ell}(\cos\theta)|^2\right]^{1/2}}
+ \mathcal{O}(\varepsilon), 
\end{equation}
with $\mu$ having the following dependence on $\varepsilon$:
\begin{equation}\label{eigenvalue}
\mu(\varepsilon) = \ell(\ell + 1) + \varepsilon \frac{\int_0^{\pi} 
d\theta\sin\theta|P_{\ell}(\cos\theta)|^4}
{\left[\int_0^{\pi}d\theta\sin\theta|P_{\ell}(\cos \theta)|^2\right]^2 } 
+ \mathcal{O}(\varepsilon^2).	
\end{equation}
For $\ell = 0$, we have the trivial constant solution 
\begin{equation}
\label{SPconstant}
f(\theta) = \frac{1}{\sqrt{2}}, 
\quad \mu(\varepsilon) = \frac{\varepsilon}{2}.
\end{equation}
For $\ell = 1$, we have a dark-soliton solution, derived from 
\eqref{modes} and \eqref{eigenvalue}, for a small expansion in $\varepsilon$, 
as  
\begin{eqnarray}
f(\theta)&=& \sqrt{\frac{3}{2}} \cos \theta + \varepsilon f_1(\theta) +
\mathcal{O}(\varepsilon^2),\label{SPdark-soliton}\\
\mu(\varepsilon) &=& 2 + \frac{9}{10}\varepsilon + \mathcal{O}(\varepsilon^2),
\label{asympt-mu-small}
\end{eqnarray} 
where $f_1(\theta)$ is a solution of
\begin{equation}
\hspace{-0.3cm}
-\!\left[\frac{d^2}{d\theta^2} + \cot \theta \frac{d}{d\theta} + 
2\right]\!\!f_1(\theta)
=  \frac{3}{10}\sqrt{\frac{3}{2}}\cos \theta\left({3}-{5} 
\cos^2 \theta\right). 
\label{SPfirst-order-correction}
\end{equation}
Under the normalization \eqref{normalization}, 
there exist a unique solution $f_1(\theta)$ of  
\eqref{SPfirst-order-correction}. Due to the orthogonality condition, 
$\int_0^{\pi} d\theta \sin \theta \cos \theta f_1(\theta) = 0$,
which follows from \eqref{normalization}, it is 
proportional to $P_3(\cos \theta)$. Hence, Laplace equation (\ref{Laplace}) with $\ell = 3$ implies that 
\begin{eqnarray}\label{f1}
f_1(\theta) &=& 
\frac{3}{100}\sqrt{\frac{3}{2}}\cos \theta 
\left(3  -5\cos^2\theta \right).
\end{eqnarray}
By substituting in \eqref{SPdark-soliton}, the solution for small 
$\varepsilon$ is
\begin{eqnarray}
f(\theta) = 
\sqrt{\frac{3}{2}}\cos\theta\left[ 1 + \varepsilon\!
\frac{3}{100} \!\left( 3- 5\cos^2\theta\right)
+  \mathcal{O}(\varepsilon^2)\right], 
\hspace{1cm} \label{SPdark-soliton-small}
\end{eqnarray}
which obeys \eqref{normalization} up to $\mathcal{O}(\varepsilon^2)$.

\section{Dark solitons for large $\varepsilon$}
We recall that the dark soliton profile $f(\theta)$ vanishes at 
$\theta = \frac{\pi}{2}$. As $\varepsilon$ increases $(\varepsilon\gg 1)$, 
its reduction becomes concentrated near $\theta = \frac{\pi}{2}$. 
The asymptotic solution is 
\begin{equation}
\label{ode-hom}
-f_{\infty}''(\theta) + \varepsilon f_{\infty}^3(\theta) = \mu_{\infty}(\varepsilon) 
f_{\infty}(\theta). 
\end{equation}
By connecting it with the constant solution \eqref{SPconstant}, and redefining it as 
$f_{\infty}(\theta)\equiv g_0(z)$, 
with $z \equiv \frac{\sqrt{\varepsilon}}{2} \left(\frac{\pi}{2} - \theta \right)$,
we have $g_0(z) = \frac{1}{\sqrt{2}} \tanh(z)$ as an exact solution of \eqref{ode-hom}:
\begin{equation}
-\frac{\varepsilon}{4}g_0''(z)+\varepsilon g_0^3(z) = 
\frac{\varepsilon}{2}g_0(z) = \mu_{\infty}(\varepsilon)g_0(z),
\label{SPdark-soliton-lim}
\end{equation}
where $\frac{d}{d\theta}=-\frac{\sqrt{\varepsilon}}{2}\frac{d}{dz}$.
Further, with $f(\theta)\equiv g(z)$, the original equation \eqref{ode} can be written as
\begin{equation}
\label{ode-equiv}
-\frac{\varepsilon}{4}g''(z) +\frac{\sqrt{\varepsilon}}{2} 
\tan\left(\frac{2 z}{\sqrt{\varepsilon}}\right) g'(z) + \varepsilon g^3(z) 
= \mu g(z).
\end{equation}
A formal expansion for $\varepsilon \to \infty$ generates the term $z g'(z)$ 
at lowest order $\mathcal{O}(1/\varepsilon)$.
However, since we need to use the normalization condition \eqref{normalization},
the asymptotic expansions of $g(z)$ and $\mu(\varepsilon)$, 
for $\varepsilon \to \infty$, are modified by the $\mathcal{O}(1/\sqrt{\varepsilon})$ terms, such that 
\begin{eqnarray}
\label{SPdark-soliton-inf}
g(z) &=& g_0(z) + \frac{1}{\sqrt{\varepsilon}} g_1(z) + \mathcal{O}\left(\frac{1}{\varepsilon}\right),\\ 
\mu(\varepsilon)&=& \frac{\varepsilon}{2}\left[1+ \frac{2\mu_1}{\sqrt{\varepsilon}} + 
\mathcal{O}\left(\frac{1}{\varepsilon}\right)\right].
\label{mularge}
\end{eqnarray}
Substituting in \eqref{ode-equiv}, we obtain the equation for $g_1(z)$:
\begin{equation}
\label{SPfirst-order}
-g_1''(z) + [4- 6\; {\rm sech}^2(z)] g_1(z)= 4 \mu_1 g_0(z).
\end{equation}
Since $\mathcal{L}_+ := -\partial_z^2 + [4 - 
6 \,{\rm sech}^2(z)]$ has a kernel spanned by ${\rm sech}^2(z)$, and 
$g_0(z)=\frac{1}{\sqrt{2}}\tanh(z)$ is an odd function,
a unique bounded solution exists for \eqref{SPfirst-order}, given by
{\small\begin{equation}
\label{SPfirst-order-solution}
g_1(z) = \frac{\mu_1}{\sqrt{2}}\left[ \tanh(z) + z \,{\rm sech}^2(z) \right]=\mu_1\frac{d}{dz}[zg_0(z)].
\end{equation}
}Next, we can fix $\mu_1$ using the
normalization constraint \eqref{normalization}, together with 
\eqref{SPdark-soliton-inf} and \eqref{mularge}. From the expansion
\begin{eqnarray}\label{SPfirst-order-2}
g^2(z) &=& g_0^2(z) + \frac{2}{\sqrt{\varepsilon}}g_0(z)g_1(z) +
\mathcal{O}\left(\frac{1}{\varepsilon} \right) 
\\
&=& g_0^2(z)\left[1+ \frac{2\mu_1}{\sqrt{\varepsilon}}\right]+ 
\frac{\mu_1}{\sqrt{\varepsilon}}z\frac{d}{dz}g_0^2(z) +
\mathcal{O}\left(\frac{1}{\varepsilon}\right),\nonumber
\end{eqnarray}
and observing that $g_0(z)=f_\infty(\theta)$ is normalized to one only asymptotically,
for $\sqrt{\varepsilon}\to\infty$, such that  
\begin{eqnarray}
& \quad & \int_0^{\pi}d \theta \sin \theta\; g_0^2(z)\Big|_{z=\frac{\sqrt{\varepsilon}}{2} \left(\frac{\pi}{2} 
- \theta \right)} \nonumber \\ &=& 1-\frac{2}{\sqrt{\varepsilon}}+\mathcal{O}\left(\frac{1}{\sqrt{\varepsilon^{3}}}\right),
\end{eqnarray} 
the normalization constraint \eqref{normalization} yields 
\begin{eqnarray}
1&=&\int_0^{\pi}d \theta \sin \theta f^2(\theta)=\frac{2}{\sqrt{\varepsilon}}  
\int_{-\frac{\pi \sqrt{\varepsilon}}{4}}^{\frac{\pi \sqrt{\varepsilon}}{4}}dz 
\cos\left(\frac{2z}{\sqrt{\varepsilon}} \right) g^2(z) \nonumber \\
&=&\left(1+ \frac{2\mu_1}{\sqrt{\varepsilon}}\right)\left(1-\frac{2}{\sqrt{\varepsilon}}\right)
+ \mathcal{O}\left(\frac{1}{\varepsilon} \right)\nonumber\\
&+&\frac{2\mu_1}{\varepsilon}
\int_{-\frac{\pi\sqrt{\varepsilon}}{4}}^{\frac{\pi\sqrt{\varepsilon}}{4}}dz 
\cos\left(\frac{2z}{\sqrt{\varepsilon}} \right){z}\frac{d}{dz} 
g_0^2(z) \nonumber \\
&=& 1 + \frac{2 (\mu_1-1)}{\sqrt{\varepsilon}}
+ \mathcal{O}\left(\frac{1}{\varepsilon} \right).
\label{norm}
\end{eqnarray}
Hence, 
$\mu_1 = 1$ in \eqref{SPdark-soliton-inf}, to satisfy 
\eqref{norm} up to the $\mathcal{O}(1/\varepsilon)$ order, implying
\begin{equation}
\label{asympt-mu-large}
\mu(\varepsilon) = \frac{\varepsilon}{2}\left[1 + \frac{2}{\sqrt{\varepsilon}} + 
\mathcal{O}\left(\frac{1}{\varepsilon} \right)\right], \quad 
\mbox{\rm as} \;\; \varepsilon \to \infty.
\end{equation}

\section{Stability analysis}
For the stability analysis, the spectrum for each $m$-angular mode is considered 
separately, with the following coupled system for the operators $L_m^\pm$:
\begin{equation}\label{SPnls-spectrum}
\left\{ \begin{array}{l} 
\omega \hat{u}_m = L^-_m \hat{v}_m, \quad L^-_m = -\Delta_m +\varepsilon f^2(\theta)  - \mu,  \\
\omega \hat{v}_m = L^+_m \hat{u}_m, \quad L^+_m = -\Delta_m +3\varepsilon f^2(\theta) - \mu.
\end{array} 
\right.
\end{equation}
where 
$$
\Delta_m = \frac{d^2}{d\theta^2} + 
\cot \theta\;\frac{d}{d \theta} - \frac{m^2}{\sin^2\theta}.
$$
We need to determine the number of negative and zero eigenvalues of the 
operators $L^{\pm}_m$, if $\mu$ and $f(\theta)$ are defined along the branch of dark 
soliton solutions \eqref{SPdark-soliton}. 
If $L^{\pm}_m$ are strictly positive, then the respective
eigenvalues $\omega_m^{\pm}$ in the spectral stability problem \eqref{SPnls-spectrum} are real 
\cite{SMGP25}, implying that the dark solitons are spectrally stable with respect to 
the $m-$angular Fourier mode. 
The number of unstable eigenvalues $\omega$ with ${\rm Im}(\omega) \neq 0$ can be 
controlled by the number of negative eigenvalues of $L_m^{\pm}$ and the
multiplicity of their zero eigenvalues (see Theorems 1.7, 1.8, and 3.10 in \cite{SMGP25}). \\

About the eigenvalues of $L^{\pm}_m$, the following facts are applied for every 
integer $m \geq 0$:
\begin{itemize}[leftmargin=10pt,labelsep=*,align=left]
\item They are simple because there may be at most one bounded solution of 
the second-order differential equation $L^{\pm}_m \chi_m^\pm(\theta) = 
\omega_m^\pm \chi_m^\pm(\theta)$ at each endpoint of the interval $[0,\pi]$.
The eigenfunctions $\chi_m^\pm(\theta)$ in the domain of $L_m^{\pm}$ provide the 
connections between the bounded solutions as $\theta \to 0$ and $ \to \pi$. 
\item The eigenfunctions are either even or odd with respect to the midpoint 
$\theta = \frac{\pi}{2}$, since $f^2(\theta)$ is even about $\theta = \frac{\pi}{2}$.
Hence, if the eigenvalues are simple, 
then the normalized eigenfunctions satisfy 
$\chi_m^\pm(\pi - \theta) = \pm \chi_m^\pm(\theta)$ (plus sign for even and
minus sign for odd functions).	
\item The smallest eigenvalue of $L_m^{\pm}$ is associated with the even eigenfunction 
about $\theta = \frac{\pi}{2}$, with the second smallest 
associated with the odd eigenfunction about $\theta = \frac{\pi}{2}$. 
\end{itemize}
The operators $L^\pm_m$  enjoy two comparative relations:
\begin{eqnarray}
\label{SPcompar-1}
L_m^+ - L_m^- = 2 \varepsilon f^2(\theta) \geq 0,\\ 
\label{SPcompar-2}
L_{m+1}^{\pm} - L_m^{\pm} = \frac{2m+1}{\sin^2\theta} \geq 0.
\end{eqnarray}
Since $L_m^+ > L_m^-$ for all $\theta \in [0,\pi]\backslash\{\frac{\pi}{2}\}$ 
and $L_{m+1}^{\pm} > L_m^{\pm}$ for all $\theta \in (0,\pi)$, 
the smallest eigenvalue of $L_m^-$ is smaller than the smallest one of $L_m^+$; 
and the smallest eigenvalue of $L_m^{\pm}$ is smaller than the smallest one of
$L_{m+1}^{\pm}$. The same holds for the second smallest 
eigenvalues of the same operators in the subspace of odd functions about 
$\theta = \frac{\pi}{2}$. 
The main results for the modes $m = 0$, $m = 1$, and $m \geq 2$ are presented below. 
\begin{description}[leftmargin=10pt,labelsep=*,align=left]
\item[\underline{Mode $m=0$}]  
We prove for the spectral problem \eqref{SPnls-spectrum} 
that there exists a single pair of eigenvalues $\omega$ of negative energy which are smaller 
than all other pairs of eigenvalues $\omega$ 
of positive energy for small values of $\varepsilon$. These pairs can coalesce hypothetically 
for large values of $\varepsilon$, 
triggering another instability bifurcation. However, our numerical data does not support 
evidence that this instability can occur for $m = 0$.

\begin{itemize}[leftmargin=10pt,labelsep=*,align=left]
	\item There exists a simple zero eigenvalue of $L_0^-$ with the eigenfunction given 
    by the profile $f(\theta)$ of the dark soliton because  $L^-_0 f(\theta) = 0$ is equivalent 
    to \eqref{ode} for every $\varepsilon > 0$. 
	
	\item By Sturm's nodal theory, $L^-_0$ has a simple negative eigenvalue since $f(\theta)$ 
    has a single node on $[0,\pi]$. The rest of the spectrum of $L_0^-$ is strictly positive 
    for every $\varepsilon > 0$.  
	
	\item Since the smallest eigenvalue of $L^-_0$ in the space of odd functions about 
    $\theta = \frac{\pi}{2}$ is located at $0$, the comparison  \eqref{SPcompar-1} implies 
    that $L^+_0$ has at most one negative eigenvalue and the second eigenvalue of $L_0^+$ 
    is strictly positive for every  $\varepsilon > 0$.  
	
	\item Since the smallest eigenvalue of $L_1^+$ is $0$ for every $\varepsilon > 0$ 
    (see the case $m = 1$), the comparison \eqref{SPcompar-2} for $m = 0$ implies that $L^+_0$ 
    has exactly one negative eigenvalue for every $\varepsilon > 0$. See the left panel of 
    Fig.~\ref{SMfig01} for illustration.
	
	\item For $\varepsilon = 0$, we have $L_m^{\pm} |_{\varepsilon = 0} = -\Delta_m - 2$. 
    By using the eigenvalues
    of the Laplace equation with $m = 0$, we obtain the asymptotic approximations of eigenvalues 
    in the stability problem \eqref{SPnls-spectrum}. For small values of $\varepsilon > 0$, 
    a special pair of simple eigenvalues exists,
\begin{equation}
\pm \omega_0 = \pm 2 + \mathcal{O}(\varepsilon), 
\end{equation}
which has  negative energy, since we have for the eigenvector $(\hat{u},\hat{v})$ 
of \eqref{SPnls-spectrum} with $\pm \omega_0$:
\begin{equation}
\langle L_0^+ \hat{u}, \hat{u} \rangle = \langle L_0^- \hat{v}, \hat{v} \rangle < 0.
\end{equation}
In addition, there exists the double-zero eigenvalue due to the rotational invariance 
and a countable sequence of pairs of simple eigenvalues 
\begin{equation}
\label{eigenvalue-positive}
\pm \omega_{\ell} = \pm [\ell (\ell+1) - 2] + \mathcal{O}(\varepsilon), \quad \ell \geq 2,
\end{equation}
with the positive energies since we have for the eigenvector $(\hat{u},\hat{v})$ of 
\eqref{SPnls-spectrum} with $\pm \omega_{\ell}$:
\begin{equation}
\langle L_0^+ \hat{u}, \hat{u} \rangle = \langle L_0^- \hat{v}, \hat{v} \rangle > 0.
\end{equation}

\item Since $\omega_{\ell} > \omega_0$ for every $\ell \geq 2$, there is some 
$\varepsilon_0 > 0$ (which might be infinite) such that 
the stability problem \eqref{SPnls-spectrum} for $m = 0$ and $\varepsilon \in 
(0,\varepsilon_0)$ admits only real eigenvalues $\omega$. 

\item The only way to get complex unstable eigenvalues $\omega$ of the stability 
problem \eqref{SPnls-spectrum} for $\varepsilon \geqq \varepsilon_0$ is if eigenvalues 
$\pm \omega_0$ of negative energy coalesce with eigenvalues 
$\{ \pm \omega_{\ell}\}_{\ell = 2}^{\infty}$ of positive energy for some 
$\varepsilon = \varepsilon_0$. Numerical evidence shows that this does not 
occur and $\varepsilon_0 = \infty$.
\end{itemize}
\begin{figure}[htb!]
\centerline{
\includegraphics[width=0.5\textwidth]{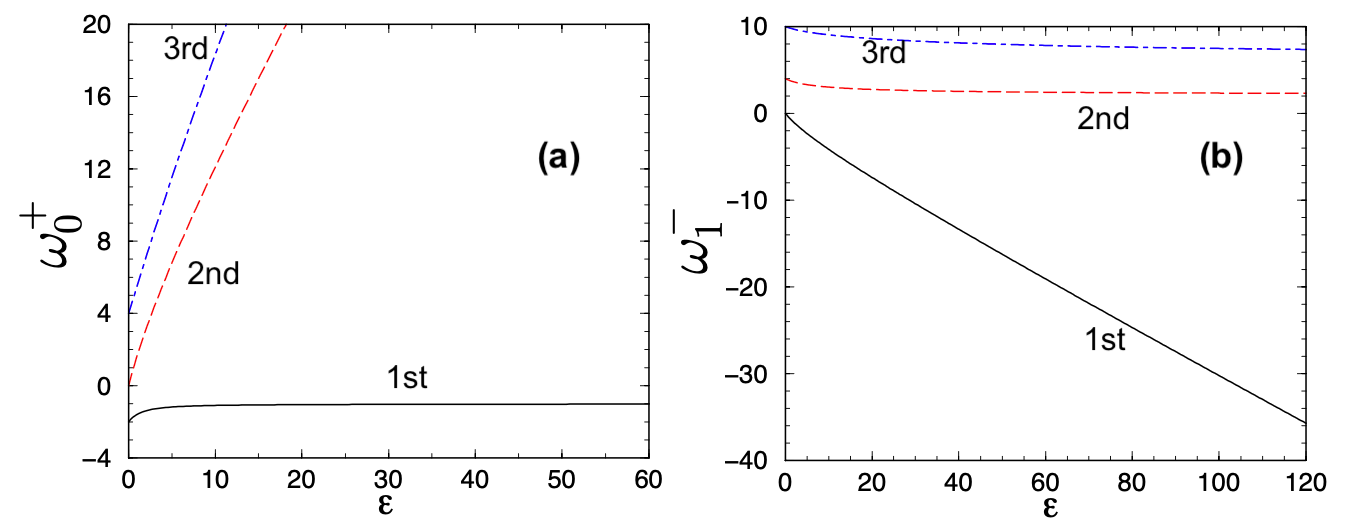} }
\caption{The three smallest eigenvalues of $L_0^+$ and $L_1^-$, 
respectively, $\omega_0^+$ and $\omega_1^-$, are given in (a) and (b), 
as functions of $\varepsilon$. The eigenvalues were 
calculated numerically by discretizing the operators with finite 
differences up to 800 points.}
	\label{SMfig01}\end{figure}

\item[\underline{Mode $m=1$}] We prove for the spectral problem \eqref{SPnls-spectrum} that only 
real eigenvalues $\omega$ exist for every $\varepsilon > 0$.
\begin{itemize}[leftmargin=10pt,labelsep=*,align=left]
\item A simple zero eigenvalue of $L_1^+$ exists, with the eigenfunction $f'(\theta)$, as
$L_1^+ f'(\theta) = 0$ is equivalent to differentiating \eqref{ode} 
    with respect to $\theta$ for every $\varepsilon > 0$.
	
\item From Sturm's nodal theory, $L^+_1$ does not have negative eigenvalues. 
$f'(\theta) < 0$ is sign-definite for $\theta \in (0,\pi)$ due to monotonicity
of the dark soliton profile \eqref{SPdark-soliton}. So, $L_1^+$ admits a simple 
zero eigenvalue, with the rest of its spectrum being strictly positive for all $\varepsilon > 0$.  
	
	\item The comparison \eqref{SPcompar-1} implies that $L^-_1$ has at least one 
    negative eigenvalue for every $\varepsilon > 0$. 
	
	\item Since the smallest eigenvalue of $L_0^-$ in the space of odd functions 
    about $\theta = \frac{\pi}{2}$ is located at $0$ for every $\varepsilon > 0$ 
    (see the case $m = 0$), the comparison \eqref{SPcompar-2} for $m = 0$ implies 
    that $L^-_1$ has exactly one negative eigenvalue and the second eigenvalue of
    $L_1^-$  is strictly positive for every $\varepsilon > 0$. See the right panel of 
    Fig.~\ref{SMfig01} for illustration.
	
	\item For $\varepsilon = 0$, we have $L_1^{\pm} |_{\varepsilon = 0} = -\Delta_1 - 2$. 
    By using the eigenvalues of the Laplace equation for $m = 1$, we obtain the asymptotic 
    approximations of the eigenvalues in the stability 
    problem \eqref{SPnls-spectrum} for $m = 1$ and small values of $\varepsilon > 0$. 
    There exists a double-zero eigenvalue due to the derivative mode 
    $f'(\theta)$ and a 
    countable sequence of pairs of simple eigenvalues 
    $\{ \omega_{\ell} \}_{\ell = 2}^{\infty}$, defined by exactly the
    same formula \eqref{eigenvalue-positive}. 
	
	\item Since all eigenvalues have positive energy for every $\ell \geq 2$ 
	and no change of eigenvalues of $L^{\pm}_1$ occurs in $\varepsilon$, 
	the stability problem \eqref{SPnls-spectrum} for $m = 1$ admits only 
    real eigenvalues $\omega$ for every $\varepsilon > 0$.
\end{itemize}
\item[\underline{Modes $m \geq 2$}]  For these cases, it is demonstrated that all 
eigenvalues of $L_m^{\pm}$ are strictly positive for small $\varepsilon > 0$ 
and that the eigenvalues of 
$L_m^+$ remain strictly positive for every $\varepsilon > 0$. The smallest 
eigenvalue of $L_m^-$ can cross $0$ at $\varepsilon = \varepsilon_m > 0$ 
and become negative for $\varepsilon > \varepsilon_m$. If this happens, 
the smallest eigenvalue of $L_{m+1}^-$ crosses $0$ 
for larger values of $\varepsilon$ compared to the smallest eigenvalue 
of $L_m^-$, that is, $\varepsilon_m < \varepsilon_{m+1}$. Since we show 
that $\varepsilon_m = 4 m(m-1) + \mathcal{O}(1)$ as $m \to \infty$ 
in \eqref{SPasympt-eps-large}, this implies that the crossing at $\varepsilon_m$ 
exists for every $m \geq 2$, triggering instability with exactly one unstable eigenvalue 
$\omega$ in the spectral stability problem \eqref{SPnls-spectrum}. 
These analytical predictions are well illustrated by the numerical results we 
are presenting in Fig. 2 of the main text, for angular modes  $m=$2, 3, 4, and 5. 

\begin{itemize}[leftmargin=10pt,labelsep=*,align=left]
	\item All eigenvalues of $L^{\pm}_m$ are strictly positive for small 
    $\varepsilon > 0$ since $L_m^{\pm} |_{\varepsilon = 0} = -\Delta_m - 2$ 
	and the eigenvalues of the Laplace equation 
    for $m \geq 2$ yield $\ell (\ell+1) - 2 > 0$ for $\ell \geq m$. Hence, there 
    exists $\varepsilon_0 > 0$ such that the stability problem \eqref{SPnls-spectrum} 
    for $m \geq 2$ and $\varepsilon \in (0,\varepsilon_0)$ admits only pairs of 
    simple real eigenvalues $\{\omega_{\ell} \}_{\ell = m}^{\infty}$ of positive 
    energy, defined by the same formula \eqref{eigenvalue-positive}.
	
\item Since the smallest eigenvalue of $L_1^+$ is $0$ for every $\varepsilon > 0$ (see the case $m = 1$), the comparison \eqref{SPcompar-2} for $m \geq 1$ 
    implies that the smallest eigenvalue of $L^+_m$ for $m \geq 2$ is strictly
    positive for every $\varepsilon > 0$.
	
\item Since the smallest eigenvalue of $L_1^-$ in the space of odd functions 
    about $\theta = \frac{\pi}{2}$ is strictly positive for every $\varepsilon > 0$
    (see the case $m = 1$), the comparison \eqref{SPcompar-2} with $m \geq 1$ 
    implies that the second eigenvalue of $L_m^-$ for $m \geq 2$ is strictly
    positive for every $\varepsilon > 0$.
		
\item The comparison \eqref{SPcompar-1} implies that the smallest eigenvalue for
$L^-_m$ is always smaller than the smallest eigenvalue for $L^+_m$. Therefore, it can be both positive and negative. The complex eigenvalues $\omega$ in the spectral stability problem 
    \eqref{SPnls-spectrum} may arise if and only if the smallest positive 
    eigenvalue of $L^-_m$ crosses $0$ at $\varepsilon = \varepsilon_m$. 
    The comparison \eqref{SPcompar-2} for $m \geq 2$ implies that the smallest eigenvalue 
    of $L^-_m$ always crosses $0$ for smaller values of $\varepsilon$ compared 
    to the smallest eigenvalue of $L^-_{m+1}$ so that $\varepsilon_m < 
    \varepsilon_{m+1}$ for $m \geq 2$.
\end{itemize}
\end{description}

\section{Asymptotic formula for $\varepsilon_m$ as $m \to \infty$}
\noindent By using the asymptotic solution \eqref{mularge} 
with $\mu_1 = 1$, together with \eqref{SPfirst-order-2}, and assuming 
$m^2 = \mathcal{O}(\varepsilon)$ as $\varepsilon \to \infty$,
{\small\begin{eqnarray}
L^-_m &=& 
-\frac{\varepsilon}{4}\partial_z^2 +  \tan\left( \frac{2z}{\sqrt{\varepsilon}}\right) 
\frac{\sqrt{\varepsilon}}{2}\partial_z
+ \frac{m^2}{\cos^2\left(\frac{2z}{\sqrt{\varepsilon}}\right)} 
+ \varepsilon g^2(z) - \mu \nonumber\\
&=& -\frac{\varepsilon}{4} \Big\{\partial_z^2 - \frac{4 m^2}{\varepsilon}
+ 2 {\rm sech}^2(z)\left[1+\frac{2 - 2z\tanh(z)}{\sqrt{\varepsilon}} 
\right] \nonumber\\ && + \mathcal{O}\left({\frac{1}{\varepsilon}}\right)\Big\}.
\end{eqnarray}
}The spectrum of $\mathcal{L}_- = -\partial_z^2 - 
2\;{\rm sech}^2(z)$ includes 
a simple negative eigenvalue at $-1$ with the eigenfunction spanned by 
${\rm sech}(z)$ and the continuous spectrum at $[0,\infty)$. By using 
perturbation theory for an isolated eigenvalue, we obtain that 
$L_m^-$ has a zero eigenvalue if and only if 
{\small\begin{eqnarray}
\frac{4 m^2}{\varepsilon} &=& 1 + \frac{4}{\sqrt{\varepsilon}} 
\frac{\int_{\mathbb{R}} (1-z \tanh(z)) {\rm sech}^4(z) dz}{\int_{\mathbb{R}} 
{\rm sech}^2(z) dz} + \mathcal{O}\left(\frac{1}{\varepsilon}\right)\nonumber\\
&=& 1 + \frac{2}{\sqrt{\varepsilon}} + 
\mathcal{O}\left(\frac{1}{\varepsilon}\right),
\end{eqnarray}
} which yields the quantization formula for bifurcations at 
$\{ \varepsilon_m \}_{m = 2}^{\infty}$. From the above,  
{\small $4m^2=\left(\sqrt{\varepsilon_m}+1\right)^2-1+\mathcal{O}(1)$}, 
\begin{eqnarray} 
&&\sqrt{\varepsilon_m} + 1= \sqrt{{4m^2}+1 -\mathcal{O}(1)},\nonumber
\end{eqnarray}
which leads to
\begin{equation}
\label{SPasympt-eps-large}
\varepsilon_m \approx 4 m (m-1) + \mathcal{O}(1), \quad \mbox{\rm as} \;\; m \to \infty.
\end{equation}
In view of the inequality 
$0 < \varepsilon_m < \varepsilon_{m+1}$ for $m \geq 2$, 
the asymptotic formula \eqref{SPasympt-eps-large} implies that the 
lowest eigenvalue of $L_m^-$ crosses $0$ at $\varepsilon = 
\varepsilon_m$ for every $m \geq 2$. 

\section{Numerical method for dark solitons}

\noindent We define  $\tilde{f}\equiv\tilde{f}(\theta) =f(\theta) \sqrt{\varepsilon}$ for solutions 
of \eqref{ode}, 
\begin{equation}
\mu \tilde{f}=-\frac{d^2\tilde{f}}{d\theta^2} - \cot(\theta) 
\frac{d\tilde{f}}{d\theta} + \tilde{f}^3,
\label{a2}
\end{equation}
subject to the normalization 
\begin{equation}
\int_0^\pi d\theta \cos(\theta) |\tilde{f}(\theta)|^2 = \varepsilon.
\label{a3}
\end{equation} 
The problem is reformulated as for a given $\mu$, we obtain the solution $\tilde{f}(\theta)$ 
from \eqref{a2} and define $\varepsilon$ from \eqref{a3}. 
Since we are looking for dark solitons, we solve \eqref{a2} just from $\theta = 0$ 
to $\theta = \pi/2$ and then take the odd continuation of $\tilde{f}(\theta)$ from 
$\theta = \pi/2$ to $\theta = \pi$. It is a two-point boundary-value problem 
with the boundary conditions $\tilde{f}'(0)=0$ and  $\tilde{f}(\pi/2)=0$.\\ 

\begin{figure}[ht]
	\centering
	\includegraphics[scale=0.4]{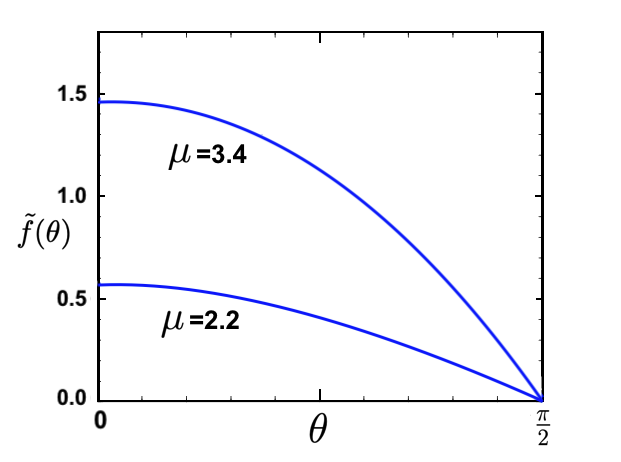}
	\caption{Shooting method for dark solitons. The left boundary condition 
    is $\tilde{f}'(0)=0$. For fixed $\mu$ one shoots $\tilde{f}(0)$ and 
    propagate Eq.~\eqref{a2} till the condition $\tilde{f}(\pi/2)=0$ is satisfied.}
	\label{SMfig02}
\end{figure}

\noindent The boundary-value problem is solved by the shooting method combined with 
the secant method \cite{SMquinney}. In this case, for a given $\mu$ we shoot 
two close values of $\tilde{f}(0)$ and propagate Eq.~\eqref{a2} with the 
Runge-Kutta method till $\theta=\pi/2$. From the values obtained of
$\tilde{f}(\pi/2)$ we can estimate a new initial shot by the secant 
method until we get $\tilde{f}(\pi/2)=0$. \\
\noindent Results for $\mu=$2.2 and 3.4 are shown in Fig.~\ref{SMfig02}. 
Sweeping $\mu$ from $2< \mu \leq 20$ can be done by continuation 
\cite{SMgammal1999}, which is performed as follows. Once we get a solution 
$\tilde{f}_\mu(0)$ for a given $\mu$, we increment $\mu$ by $\delta \mu=0.1$ 
and use $\tilde{f}_\mu(0)$ as an initial ansatz, and after shooting 
combined with the secant method, we obtain $\tilde{f}_{\mu+\delta \mu}(0)$.
Subsequently, we keep incrementing $\mu$ by $\delta \mu$ and take the 
previous $\mu$ initial value. For $\mu > 20$, continuation is not 
effective. For larger values of $\mu$, we observe from
\eqref{SPdark-soliton-lim} that a good approximation is given by 
$\tilde{f}(0)=\sqrt{\mu}$. By using this initial ansatz for each $\mu$, 
it was possible to obtain the dark soliton solutions from $\mu=2$ to $\mu = 80$.
The value of $\varepsilon$ is obtained {\it a posteriori} by using Eq.~\eqref{a3}. The numerical approximations of 
$f(\theta)$ and $\mu(\varepsilon)$ are shown in Fig. 1 of the main text.

\section{Numerical method for the GPE}

\noindent We consider the time-dependent evolution of the GPE 
for $\psi\equiv \psi(\theta,\phi,t)$, as in~\cite{SM2021Andriati}, with 
linear part given by
\begin{eqnarray}
{\rm i}\frac{\partial\psi}{\partial t}&=&-\left[
\frac{1}{\sin\theta}{\partial_\theta}\left(\sin\theta{\partial_\theta}\right)+
    \frac{1}{\sin^{2}\theta}{\partial^2_\phi}
\right]\psi
\label{nabla2}.
\end{eqnarray}

\noindent To avoid problems at $\theta=\{0,\pi\}$, the $\psi$ is expanded 
in its Fourier modes, as 
\begin{equation}
    \psi(\theta,\phi,t)=\sum_k e^{i k\phi}\psi_k(\theta,t)
    \label{fourier}.
\end{equation}
A grid of $M+1$ points is defined, with $\theta_j=jh, j\in 0,1,...M$, where $h=M/\pi$.
For each $\theta_j$, $j=1,\dots,M-1$, $\psi_k$ is computed by using a fast Fourier 
transform algorithm in the $\phi$ direction (${\cal{F}}_\phi$). In the poles, where 
$\theta=\{0,\pi\}$ and there is no dependence on $\phi$, only the mode $k=0$ 
contributes. From~\eqref{fourier} and~\eqref{nabla2}, we have the following
equation to be solved for $\psi_k\equiv\psi_k(\theta,t)$, 
in the interval $\theta \in (0,\pi)$ with boundary conditions 
$\psi_k(0,t)=\psi_k(\pi,t)=0, \forall k\neq 0 $:
\begin{equation}{\rm i}\frac{\partial \psi_k}{\partial t}=-\left[\frac{\partial^2\psi_k}
    {\partial \theta^2}+\cot\theta \frac{\partial \psi_k}{\partial \theta}
    -\frac{k^2}{\sin^2\theta}\psi_k
    \right],\end{equation} 
Equation~\eqref{nabla2} is evolved one time step $\delta t$ using the finite difference
Crank-Nicolson method for each $k$,  (${\rm CN}_{\theta,\delta t, k}$).
For $k=0$, it is required the Neumann boundary conditions, 
${\partial \psi_0}/{\partial \theta} \large |_{0,\pi}=0$. We added 
one extra point at each boundary to implement these conditions. 
After the evolution, we perform the inverse Fourier transform in the $\phi$ 
direction.  To include the nonlinear term $g|\psi|^2$, we employ the split-step 
operator technique. The complete scheme for one time step evolution is given by
\begin{equation}
\psi(\theta,\!\phi,\!t+\delta t)\!
=\!\!e^{-\frac{{\rm i} g|\psi|^2 \delta t}{2}}\!
{\cal{F}}_{\phi}^{-1}
{\rm CN}_{\theta, \delta t,k}
{\cal{F}}_\phi{\psi(\theta,\!\phi,\! t)}e^{-\frac{{\rm i} g|\psi|^2 \delta t}{2}},
\end{equation}
where the computational operations are performed in a sequence from right to left. The numerical solutions of \eqref{nabla2} are shown in Fig. 3 of the main text.

\end{document}